\pdfoutput=1
\RequirePackage{rotating}
\documentclass[fleqn,usenatbib]{mnras}  

\usepackage[T1]{fontenc}
\usepackage{ae,aecompl}

\usepackage{newtxtext}
\usepackage[varvw]{newtxmath}

\usepackage{graphicx}	
\usepackage{amsmath}	
\usepackage{amssymb}	
\usepackage[british]{babel}
\usepackage[]{units}
\usepackage{booktabs}
\usepackage{xcolor}
\usepackage{microtype}
\hypersetup{pdfauthor={Errani et al.},
            pdftitle={},
            pdfkeywords={Local Group, galaxies: kinematics and dynamics, galaxies: dwarf, dark matter},
            bookmarksnumbered=true}

\newcommand{\rmax}{r_\mathrm{max}}	%
\newcommand{\Mmax}{M_\mathrm{max}}	%
\newcommand{\kms}{\mathrm{km\,s^{-1}}}		%
\newcommand{\Rh}{R_\mathrm{h}}	%
\newcommand{\declog}{\mathrm{log_{10}}}	%
\newcommand{\DF}{\textsc{df}}	%
\newcommand{\diff}{\mathrm{d}}
\newcommand{\Msol}{\mathrm{M_{\odot}}}
\newcommand{\tc}{t_\mathrm{c}}

\title[Can tides disrupt cold dark matter subhaloes?]{Can tides disrupt cold dark matter subhaloes?}

\author[Errani \& Pe\~narrubia]{Rapha\"el Errani\thanks{E-mail: raer@roe.ac.uk} \& Jorge Pe\~narrubia
\\
{Institute for Astronomy, University of Edinburgh, Royal Observatory, Blackford Hill, Edinburgh EH9 3HJ, UK}\\
}

\date{Accepted 2019 November 27. Received 2019 November 24; in original form 2019 May 30.}

\pubyear{2019}
\volume{}

\begin{document}

\label{firstpage}
\pagerange{\pageref{firstpage}--\pageref{lastpage}} \pubyear{2019}
\maketitle

\begin{abstract}
The clumpiness of dark matter on sub-kpc scales is highly sensitive to the tidal evolution and survival of subhaloes. In agreement with previous studies, we show that $N$-body realisations of cold dark matter subhaloes with centrally-divergent density cusps form artificial constant-density cores on the scale of the resolution limit of the simulation. These density cores drive the artificial tidal disruption of subhaloes. We run controlled simulations of the tidal evolution of a single subhalo where we repeatedly reconstruct the density cusp, preventing artificial disruption. This allows us to follow the evolution of the subhalo for arbitrarily large fractions of tidally stripped mass. Based on this numerical evidence in combination with simple dynamical arguments, we argue that cuspy dark matter subhaloes cannot be completely disrupted by smooth tidal fields.
Modelling stars as collisionless tracers of the underlying potential, we furthermore study the tidal evolution of Milky Way dwarf spheroidal galaxies. Using a model of the Tucana III dwarf as an example, we show that tides can strip dwarf galaxies down to sub-solar luminosities. The remnant \emph{micro-galaxies} would appear as co-moving groups of metal-poor, low-mass stars of similar age, embedded in sub-kpc dark matter subhaloes. 
\end{abstract}



\section{Introduction}

The hierarchical clustering of dark matter (DM) is a remarkably successful framework to explain structure formation on large galactic scales.
However on kpc scales and smaller, the clustering properties of DM are subject of controversy and debate. It is on these scales that potential DM particle properties \citep[e.g.][]{TremaineGunn1979,Vogelsberger2012} as well as baryonic effects \citep[e.g.][]{Navarro96,Pontzen2012,Read2016cores} leave their imprints on the DM distribution. 
While DM-only cosmological simulations predict a universal density profile with a cusp of central slope $\gamma = - \diff \ln \rho / \diff \ln r = 1$ \citep{nfw1997}, kinematic studies of stars in DM dominated Milky Way dwarf galaxies did not yield conclusive evidence whether potential underlying DM profiles have centrally-divergent density cusps \citep[e.g.][]{RichardsonFairbairn2014}, or constant-density cores \citep[e.g.][]{Walker2011,AmoriscoAgnelloEvans2013}. 
The number of known Milky Way dwarf galaxies has increased dramatically over recent years, with deep photometric and kinematic surveys revealing subsequently fainter and less massive satellites \citep[e.g.][]{Drlica-Wagner2015,Koposov2015,Torrealba2016}. Nevertheless, their abundance can be matched to the vast number of subhaloes in cosmological simulations only by either facilitating the tidal disruption of subhaloes before redshift $z=0$, or by suppressing star formation in low-mass subhaloes. This can be achieved by involving baryonic processes \citep[e.g.][]{Chan2015Fire, SchayeEagle2015, Despali2017}, or DM recipes departing from the classical cold dark matter (CDM) model \citep[e.g.][]{Lovell2014}. Several methods have been proposed in recent years to detect also subhaloes devoid of stars, indirectly through their effects on tidal streams \citep[e.g.][]{Ibata2002heating, ErkalBelokurov2014}, or more directly through strong gravitational lensing \citep[e.g.][]{Vegetti2009} -- though clear signatures of such dark subhaloes are yet to be discovered.

The presence of self-bound subhaloes within larger DM haloes as relics of their accretion history was noted as soon as cosmological simulations had sufficient resolution to probe the scales in question \citep[resolving main haloes with $\sim10^4$ particles, e.g.][]{Tormen1997, Moore1998}. It was soon understood that insufficient resolution depletes subhaloes artificially \citep[which was originally dubbed \emph{overmerging}, e.g.][]{Klypin1999}. 
Current DM-only simulations of Milky Way-like haloes resolve DM subhaloes of masses down to $\sim \unit[10^5]{\Msol}$ at an $N$-body particle mass of $\unit[10^3 - 10^4]{\Msol}$ 
\citep[resolving main haloes with $10^8 - 10^9$ particles, e.g.][]{Springel2008, Griffen2016}. Recent studies raise suspicion whether the predictions on abundance and structural parameters of subhaloes at these mass scales can be trusted: \cite{vdb2018} argues that up to 80 per cent of subhaloes that disrupt in cosmological simulations do so because of numerical issues. This is also supported by the results of controlled simulations which suggest that DM  subhaloes with centrally-divergent density cusps cannot be fully disrupted by tides \citep{Kazantzidis2004,Goerdt2007,Penarrubia2010,vdBOgiya2018} -- although also in these simulations, subhaloes do disrupt eventually due to limited resolution and finite particle number. 

In this paper, we study the tidal evolution of a single cuspy DM subhalo under the assumption that tides do not alter the central slope of $\gamma = 1$, as suggested by the results of controlled simulations \citep{Hayashi2003,Penarrubia2010}. Our choice of $\gamma = 1$ is motivated by the \citet{nfw1997} density profile for DM haloes. While other authors find slightly steeper \citep[][for subhaloes]{Diemand2008Nature} or slightly shallower slopes \citep[e.g.][]{NavarroLudlow2010, LudlowNavarro2013}, within the resolution limits, density profiles in DM-only simulations are cuspy, i.e. centrally-divergent. We follow the tidal evolution of our example cuspy subhalo in an evolving, analytical host potential, periodically reconstructing the density cusp while simultaneously increasing the spatial resolution of the simulation. This procedure prevents artificial disruption and allows us to study the tidal evolution over arbitrarily large fractions of stripped mass.

The apparent \emph{indestructibility} of cuspy subhaloes also has implications for dwarf galaxies embedded in such haloes: As an illustration, we follow the tidal evolution of a dwarf galaxy embedded in a cuspy DM halo using an $N$-body model tailored to match the ultra-faint Tucana III dwarf galaxy \citep{Drlica-Wagner2015}. We chose the Tuc III dwarf as an example as several of its measured structural and kinematic properties indicate strong past tidal interactions: Tuc III is on a very radial orbit with a pericentre distance of $\sim\unit[3]{kpc}$, passing through the galactic disc, and has an associated stellar tidal stream \citep{LiTucana2018,Shipp2018}.
The luminosity $L\sim\unit[10^3]{L_\odot}$ and line-of-sight velocity dispersion $\sigma < \unit[1.5]{\kms}$ of the dwarf are particularly low \citep{Simon2017}, suggesting that Tuc III might be the remnant of a more massive and more luminous progenitor.
In this work, we model the tidal stripping of Tuc III down to sub-solar luminosities: Interestingly, the remnant \emph{micro-galaxy} would appear as a co-moving group of metal-poor stars of similar age, embedded in a sub-kpc DM halo.

The paper is structured as follows: In section \ref{sec:tdyn}, we present simple dynamical arguments for the distinct tidal evolution and survival of DM substructures with density cusps and cores. Following the lead of \citet{vdb2018}, we show in section \ref{sec:coreform_main} how limited resolution in numerical simulations causes the artificial formation of density cores at the centres of DM subhaloes. Section \ref{sec:reconstruct_main} details our numerical experiments of the tidal evolution of a single subhalo, where we periodically reconstruct the density cusp. 
To model the evolution of Milky Way dwarf spheroidal galaxies, in section \ref{sec:tuciii} we embed stars in a DM subhalo using a distribution function based approach and study the tidal stripping of dwarf galaxies down to sub-solar luminosities. In section \ref{sec:discuss} we summarize and discuss our findings in the context of detectability of low-mass subhaloes and highly stripped dwarf galaxies.

\section{Tidal evolution of dynamical times}
\label{sec:tdyn}

\begin{figure}
 \centering
 \includegraphics{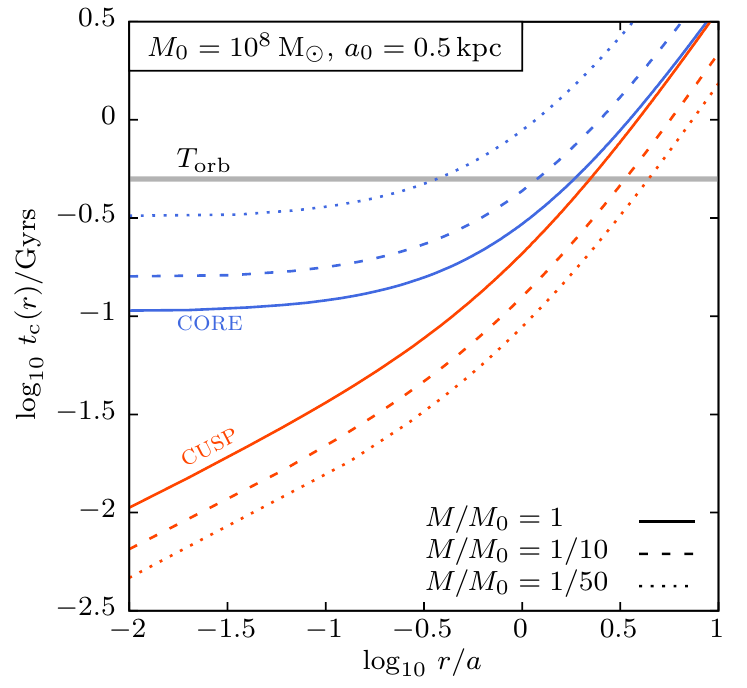}
 \caption{The internal period $t_\mathrm{c}(r)$ of a circular orbit of radius $r$ in cuspy ($\gamma=1$) \citet{dehnen1993} profiles tends towards zero for $r \rightarrow 0$, whereas for cored ($\gamma=0$) profiles, it approaches a constant, non-zero value. This is shown for subhaloes of initial mass $M_0 = \unit[10^8]{M_\odot}$ and scale radius $a_0 = \unit[0.5]{kpc}$ (solid lines).
 For cuspy (cored) profiles, tidal stripping decreases (increases) $t_\mathrm{c}(r/a)$ at fixed fractions $r/a$ of the scale radius: dashed (dotted) lines are computed using the \citet{EPW18} tidal tracks (measured from controlled simulations) and show $t_\mathrm{c}(r/a)$ when the subhalo has been tidally stripped to a remnant mass fraction of $M/M_0 = 1/10$ ($1/50$).
 As a consequence, for cuspy profiles, the fraction of particles which react adiabatically to a tidal perturbation increases with tidal stripping.
 An orbital period of $T_\mathrm{orb}=\unit[0.5]{Gyrs}$ within the host halo is shown as guidance for the time available for the subhalo to reach dynamical equilibrium between two pericentre passages, corresponding to the orbital period of the Tucana III model discussed in sections \ref{sec:reconstruct_main} and \ref{sec:tuciii}.}
 \label{fig:tcirc}
\end{figure}

Consider a subhalo on an eccentric orbit of period $T_\mathrm{orb}$ within the main halo.
Moving towards pericentre, tidal forces on the subhalo increase. 
\emph{Under which conditions does a subhalo retain some fraction of bound particles after pericentre passage?}
We address this question by contemplating the response of particles within the subhalo to the tidal field of the main halo. 
For this purpose, we compare the periods of circular orbits $\tc(r)$ within cuspy and cored subhaloes, 
and study how $\tc(r)$ evolves while the subhaloes structurally change due to tidal mass loss. We will show that for cuspy subhaloes, there is always a fraction of particles that react adiabatically to the tidal perturbation, and that this fraction increases during the tidal evolution of the subhalo.

We model the subhalo as a \citet{dehnen1993} profile with total mass $M$, scale radius $a$ and scale density $\rho_s = (3-\gamma)M/4\pi a^3$, which can be written in terms of the general $\{\alpha,\beta,\gamma\}$ profile,
\begin{equation}
\label{eq:betagammaprofile}
 \rho(r) = \rho_s \left( \frac{r}{a} \right)^{-\gamma} \left[ 1 + \left( \frac{r}{a} \right)^\alpha \right]^{(\gamma-\beta)/\alpha}~,
\end{equation}
with $\alpha=1$, outer slope $\beta\equiv-\mathrm{d}\ln \rho/\mathrm{d}\ln r~(r\rightarrow \infty)=4$ and inner slope $\gamma\equiv-\mathrm{d}\ln \rho/\mathrm{d}\ln r~(r\rightarrow 0)$. The period of a circular orbit of radius $r$ then becomes
\begin{equation}
\label{eq:tcirc}
 \tc(r) = {2 \pi}  \left[  \frac{(r+a)^{3-\gamma}}{GM\, r^{-\gamma}} \right]^{1/2} ~.
\end{equation}
Figure \ref{fig:tcirc} shows $\tc(r)$ for cuspy ($\gamma=1$) and cored ($\gamma=0$) subhaloes for an initial mass $M_0 = \unit[10^8]{M_\odot}$ and initial scale radius $a_0 = \unit[0.5]{kpc}$ (solid lines). For the cuspy model, as $r \rightarrow 0$, also $\tc(r) \rightarrow 0$, i.e. there is always a subset of radii for which $\tc(r) \ll T_\mathrm{orb}$. As the strongest tidal interaction happens on a timescale of some fraction of the orbital period $T_\mathrm{orb}$, we can assume that for particles with $\tc(r) \ll T_\mathrm{orb}$, the tidal interaction is perceived as a mere adiabatic perturbation.
Furthermore, the same particles have $T_\mathrm{orb}/\tc \gg 1$ revolutions within the subhalo to reach dynamical equilibrium before the next strong tidal interaction.  
On the other hand, for the cored model, $\tc(r) \rightarrow \mathrm{const} > 0$ as $r \rightarrow 0$: inside the density core, all orbits have the same orbital period. Whether there is a subset of particles that react adiabatically to the perturbation depends on the specific values of $\tc(r)$ and $T_\mathrm{orb}$. Similarly, the number of revolutions $T_\mathrm{orb}/\tc$ available to particles in cored subhaloes to relax before the next strong tidal interaction depends on the specific values of $\tc(r)$ and $T_\mathrm{orb}$.

\emph{How does $\tc(r)$ evolve during tidal stripping?} Subhalo mass $M$, scale radius $a$ and the shape of the density profile $\rho(r)$ all evolve due to tidal mass loss. For simplicity, in the following discussion of orbital periods $\tc(r)$ we will assume self-similar evolution of the subhalo density profile and only consider the change of subhalo mass $M$ and scale radius $a$ during tidal stripping. This assumption is well motivated for particles with $r \ll a$, as the central regions of subhaloes are shown to evolve in a self-similar manner in controlled simulations \citep{Hayashi2003,Penarrubia2010}. 
We make use of tidal evolutionary tracks \citep[originally introduced by][]{penarrubia2008} to parametrize the evolution of equilibrium halo structural parameters as a function of the fraction $M/M_0$ of remnant bound mass. In specific, we use the \citet{EPW18} tracks (measured from controlled simulations) for the evolution of the DM scale radius for cuspy and cored subhaloes. As shown with dashed (dotted) lines in Figure \ref{fig:tcirc} for remnant bound masses of $M/M_0 = 1/10$ $(1/50)$, at a fixed fraction $r/a$ relative to the instantaneous scale radius $a$, for cuspy models, the period $\tc(r/a)$ decreases during tidal stripping. Consequently, the fraction of particles in the subhalo (relative to the total instantaneous number of bound particles) for which the tidal interaction is perceived as an adiabatic perturbation increases during tidal evolution, and so does the number of revolutions $\tc(r/a)$ available for the subhalo to reach dynamical equilibrium within the (constant) orbital period $T_\mathrm{orb}$: this suggests that tides cannot fully disrupt cuspy subhaloes\footnote{Note that while for cuspy subhaloes $\tc(r/a)$ at a fixed fraction $r/a$ of the instantaneous scale radius $a$ decreases with tidal mass loss, $\tc(r)$ at a fixed value of $r$ increases. In the central regions of the cuspy subhalo however, $\tc(r)$ is only weakly effected by tidal mass loss: Assuming self-similar evolution and using the \citet{EPW18} tidal evolutionary tracks ($a \propto M^{0.48}$), equation \ref{eq:tcirc} gives $\tc(r) \propto r^{0.5} M^{-0.02}$ for $r \ll a$.}.
For cored models however, $\tc(r/a)$ increases during tidal stripping at fixed $r/a$: the region that reacts adiabatically to tides shrinks, and it becomes increasingly difficult for the subhalo to reach dynamical equilibrium within $T_\mathrm{orb}$. This drives the eventual tidal disruption of the cored subhalo. In this context, the term \emph{artificial} disruption has been coined by \citet{vdb2018} for the disruption of subhaloes in cosmological simulations caused by numerical issues, e.g. by inadequate force softening.

While the evolution of dynamical times suggests that smooth tidal fields cannot fully disrupt cuspy subhaloes, the specific rate of tidal stripping will depend on the host potential and subhalo orbit \citep[see e.g.][]{Hayashi2003}. The apparent \emph{indestructibility} of cuspy subhaloes is consistent also with their tidal radius $r_\mathrm{t}$: For a subhalo with pericentre distance $r_\mathrm{peri}$, the tidal radius can be approximated as the radius $r_\mathrm{t}$ for which $\langle \rho (<r_\mathrm{t})\rangle = 3 \langle \rho_\mathrm{host} (<r_\mathrm{peri})\rangle$ \citep[see e.g.][]{penarrubia2008}, where by $\langle \rho (<r_\mathrm{t})\rangle$ we denote the mean density of the subhalo averaged within $r_\mathrm{t}$, and equivalently by $\langle \rho_\mathrm{host} (<r_\mathrm{peri})\rangle$ the mean density of the host halo averaged within $r_\mathrm{peri}$.
The diverging central density of a cuspy subhalo, $\rho(r) \rightarrow \infty$ for $r \rightarrow 0$, guarantees the existence of a finite and non-zero tidal radius~$r_\mathrm{t}$, and therefore suggests that a fraction of particles will remain bound to the subhalo after tidal interaction.

\begin{figure*}
 \centering
 \includegraphics{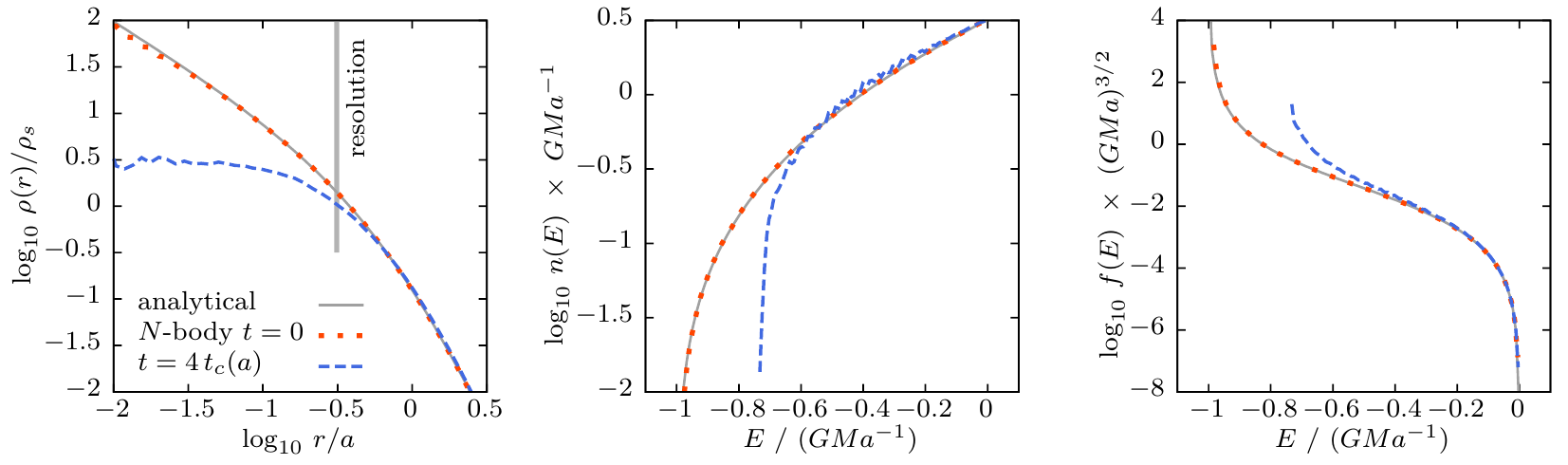}
 \caption{A cuspy \citet{dehnen1993} model ($\{\alpha,\beta, \gamma\} = \{1,4,1\}$) of mass $M$ and scale radius $a$ evolved in isolation using a particle-mesh code with a spatial resolution of $\Delta x = 20a/64 = \unit[0.16]{kpc}$ forms a density core ($\diff \ln \rho / \diff \ln r \rightarrow 0$ for $r\rightarrow 0$) on the scale of the spatial resolution of the simulation (left panel). The differential energy distribution $n(E)$ reveals that the evolved $N$-body models is missing particles with the most-negative energies compared to the analytical and unevolved model (central panel). The distribution function for less-bound energies follows closely the analytical model (right panel).} 
  \label{fig:eqcore}
\end{figure*}

\section{Core formation in numerical simulations}
\label{sec:coreform_main}
In the spirit of \citet{vdb2018}, in this section, we perform a numerical experiment to demonstrate how density cores form artificially due to insufficient spatial resolution. For this purpose, we generate an equilibrium $N$-body realisation of a cuspy DM halo (with the general method described in section \ref{sec:eqmodels}) and evolve it in isolation using the particle-mesh code \textsc{superbox} \citep{Fellhauer2000} (section \ref{sec:coreformexp}).

\subsection{Generation of equilibrium models}
\label{sec:eqmodels}
Throughout this paper, we make use of spherical equilibrium $N$-body models with isotropic velocity dispersion, and summarise in this subsection the procedure to generate such models. We aim to generate an $N$-body model of (tracer) density $\nu(r) = \diff N/\diff^3 r$ which is in dynamical equilibrium in a spherical potential $\Phi(r)$. The distribution function (hereafter \DF) $f(E) = dN/d\Omega$ which determines the number $N$ of particles of energy $E$ per phase space volume element $\diff\Omega = \diff^3 r\, \diff^3 v$ can be obtained from $\nu(r)$ and $\Phi(r)$ using Eddington inversion \citep[][]{Eddington1916}:
\begin{equation}
 f(E) = \frac{1}{\sqrt{8} \pi^2} \int_E^0 \frac{\mathrm{d}^2 \nu}{\mathrm{d}^2 \Phi} \left({\Phi-E}\right)^{-1/2}  \,\mathrm{d}\Phi ~,
 \label{eq:DF}
\end{equation}
which holds in this form if for $r \rightarrow \infty$ both $\Phi \rightarrow 0$ and \mbox{$\mathrm{d} \nu / \mathrm{d} \Phi \rightarrow 0$}.
In equation \ref{eq:DF}, the tracer density $\nu(r)$ is normalised so that $N = 4 \pi \int_0^\infty r^2 \nu(r) \diff r$.
For the case of self-gravitating models, the potential $\Phi(r)$ is sourced by the mass density $\rho(r) = m \nu(r)$, where by $m$ we denote the mass of a single particle. 
Following the notation of \citet{Spitzer1987book}, each particle of energy $E$ has access to a differential phase space volume of $\diff\Omega/\diff E = (4 \pi)^2 p(E) $, where
\begin{equation}
 p(E) = \int_0^{r_\mathrm{apo}(E)} \left\{2 \left[ E - \Phi(r) \right] \right\}^{1/2} r^2\,\mathrm{d}r ~.
 \label{eq:probE}
\end{equation}
The integration limits correspond to the minimum and maximum radii $r$ accessible to a particle with energy $E$ in the potential $\Phi(r)$.
The differential energy distribution $n(E) = \diff N / \diff E$ then becomes
\begin{equation}
n(E) = (4 \pi)^2 p(E) f(E) ~.
\label{eq:diffenergy}
\end{equation}
We deduce from equations \ref{eq:probE} and \ref{eq:diffenergy} that at given radius $r$, the likelihood of a particle to have energy $E$ is $\mathcal{L}(E|r) \propto \left\{2 \left[ E - \Phi(r) \right] \right\}^{1/2} r^2 f(E)$.
This allows us to generate equilibrium $N$-body models using a two-step procedure: we first draw radii $r$ through inverse transform sampling of the tracer density $\nu(r)$, and subsequently energies $E$ through rejection-sampling with the likelihood $\mathcal{L}(E|r)$. For isotropic systems, energies and radii uniquely determine the velocities of the $N$-body particles.
A basic implementation of this method is made available online\footnote{\url{https://github.com/rerrani}}.

\subsection{Numerical experiment for core formation}
\label{sec:coreformexp}
To illustrate the artificial formation of density cores in \emph{collisionless} $N$-body models of originally cuspy subhaloes, 
we evolve a \citet{dehnen1993} model of $N=10^7$ particles, total mass $M=\unit[10^8]{M_\odot}$ and scale radius $a=\unit[0.5]{kpc}$ in isolation using \textsc{superbox} \citep{Fellhauer2000}. This particle-mesh code employs co-moving grids centred on the densest region in the halo. We choose a low grid resolution of $\Delta x = 20a/64 = \unit[0.16]{kpc}$ for the highest-resolving grid to highlight the effect of artificial core formation. While this experiment is run in isolation, i.e. in absence of an external potential, we will show in section \ref{sec:reconstruct_sim} that the scale radius $a$ of a subhalo experiencing tidal mass loss decreases over time. Consequently, also the ratio $a/\Delta x$ decreases and can easily reach values as extreme as in our experimental setup. We chose a time step of $\Delta t = \tc(a)/400 = \unit[0.5]{Myrs}$. Note that the applied particle mesh code is collisionless and does not suffer from artificial self-heating driven by two-body relaxation.

Figure \ref{fig:eqcore} compares the analytical model and unevolved ($t=0$) collisionless $N$-body realisation against a model evolved for a time of $4 \tc(a)$. A density core forms on the scale of the resolution $\Delta x$ of the highest-resolving grid. Energies $E$, differential energy distribution $n(E)$ and \DF\ $f(E)$ are calculated for the $N$-body models using positions and velocities provided by \textsc{superbox}. We assume spherical symmetry and that both $f(E)$ and $p(E)$ are functions of energy alone, which allows $p(E)$ to be computed from equation $\ref{eq:probE}$, and $f(E) = (4 \pi)^{-2}\,n(E)\, p^{-1}(E)$, where the differential energy distribution $n(E)$ is measured directly from the $N$-body particles.
The differential energy distribution $n(E)$ of the evolved $N$-body model has fewer particles at the highest binding energies than the unevolved model and analytical counterpart, but at less-bound energies, both $n(E)$ and the \DF\ of the evolved $N$-body model follow closely the analytical form.

\begin{figure*}
 \centering
 \includegraphics{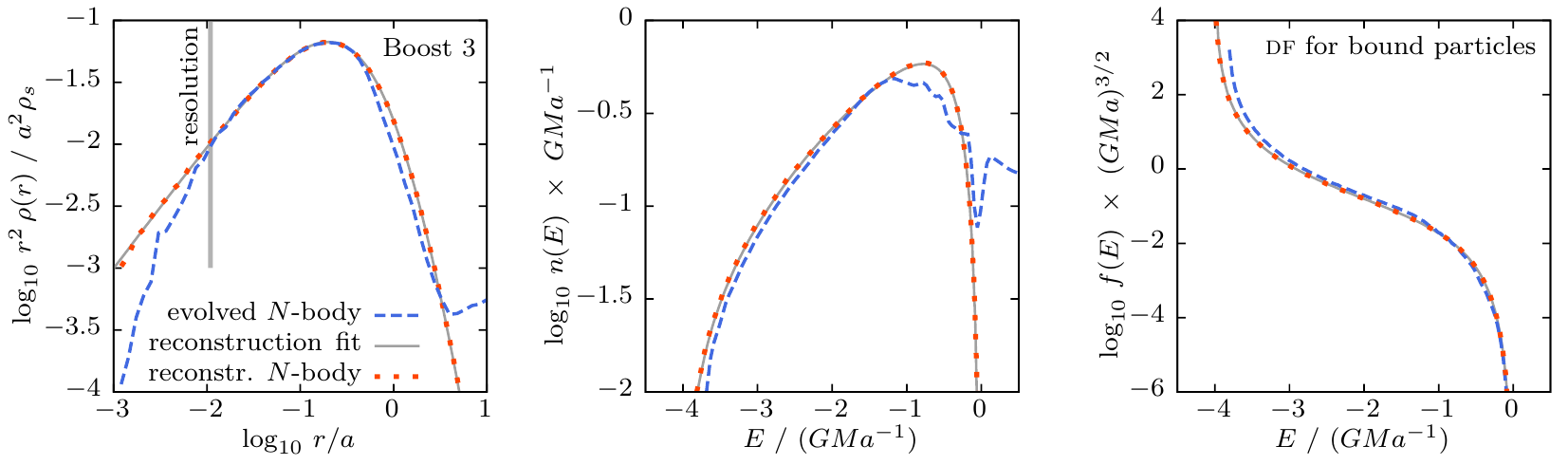}
 \caption{The left-hand panel shows an $\{\alpha,\beta,\gamma\} = \{1,7,1\}$ fit to bound $N$-body particles of a tidally stripped subhalo (\emph{evolved N-body}), reconstructing the density cusp with central slope $\gamma = 1$. The reconstructed $N$-body model corresponds to the third cusp reconstruction (\emph{boost 3}) in Figures \ref{fig:orbitplot} and \ref{fig:boostDM}, whereas the \emph{evolved N-body} model corresponds to the last apocentre snapshot of the simulation starting from the second cusp reconstruction (\emph{boost 2}).
 The fit matches the $N$-body model best for those radial shells which contain the largest number of particles, i.e. around $r_{-2}$, where the $r^2 \rho(r)$ curve peaks. The outer profile is matched with less accuracy, which can also be seen from the differential energy distribution $n(E)$ (central panel). Convergence tests however demonstrate that the tidal evolution is insensitive to an exact match of the outer density profile. }
  \label{fig:cusprec}
\end{figure*}

\section{Reconstruction of the cusp}
\label{sec:reconstruct_main}

We now explore how to reconstruct the density cusp with the aim to follow the tidal evolution of a single subhalo in a Milky Way-like host potential for arbitrarily large fractions of tidally stripped mass, avoiding artificial disruption due to insufficient numerical resolution. In section \ref{sec:numsetup} we describe the analytical, evolving host potential as well as the initial conditions for the subhalo used in our controlled simulations. The cusp reconstruction method is introduced and then applied to follow the tidal evolution of a subhalo in section \ref{sec:reconstruct_sim}. 

\subsection{Numerical setup}
\label{sec:numsetup}
\textbf{Host.} The parameters of the analytical, time-evolving host potential at redshift $z=0$ are motivated by the \citet{McMillan2011} Milky Way model with a circular velocity of $v_c = \unit[240]{km\,s^{-1}}$ at a solar radius of $R_0 = \unit[8.29]{kpc}$. We model the Milky Way disc as an axisymmetric two-component model consisting of a thin and thick \citet{miyamoto1975} disc with $M=\unit[7.3 \times 10^{10}]{M_\odot}$, $a_d = \unit[3.9]{kpc}$, $b_d = \unit[0.31]{kpc}$ ($M=\unit[2.0 \times 10^{10}]{M_\odot}$, $a_d = \unit[4.4]{kpc}$, $b_d = \unit[0.92]{kpc}$) for the thin (thick) disc, respectively. The Bulge is modelled as a \citet{hernquist1990} sphere with $M=\unit[2.1\times10^{10}]{M_\odot}$, $a=\unit[1.3]{kpc}$, and the DM halo as a spherical \citet{nfw1997} profile with scale mass $M_s=\unit[1.53 \times 10^{11}]{M_\odot}$, scale radius $r_s = \unit[20.2]{kpc}$ and concentration $c=9.49$, which results in a virial mass of $M(<cr_s) = \unit[1.40 \times 10^{12}]{M_\odot}$. The scale mass evolves with redshift, $M_s(z) \propto\exp(-2a_gz)$, whereas $r_s(z) \propto \exp(-2a_gz/\gamma_g)$, following the model by \citet[][]{buisthelmi14} with parameters $\gamma_g=2$ for strict inside-out growth and $a_g = 0.2$, motivated as a rough mean of the values found for the Aquarius \citep{Springel2008} simulations. 
We use the same recipe for the evolution of disc and bulge.
As cosmology, we adopt $\Omega_\mathrm{m} = 0.32$, $\Omega_\mathrm{\Lambda} = 0.68$, $H_0 = \unit[67]{km\,s^{-1}\,Mpc^{-1}}$ \citep{Planck2018}.

\noindent\textbf{Subhalo.} We model the subhalo at infall as an equilibrium \citet{dehnen1993} profile with $10^8$ particles, total mass $M=\unit[10^8]{M_\odot}$ and scale radius $a = \unit[0.5]{kpc}$ using the method described in section \ref{sec:eqmodels}. These structural parameters are chosen to be compatible with a progenitor of the ultra-faint Tucana III dwarf galaxy, as will be detailed in section \ref{sec:tuciii}, and correspond to a maximum circular velocity $v_\mathrm{max} = \unit[15]{km\,s^{-1}} $ at a radius $r_\mathrm{max} = a = \unit[0.5]{kpc}$. These values lie within one standard deviation from the mean $r_\mathrm{max}$-$v_\mathrm{max}$ relation for subhaloes found in the Aquarius simulations \citep[][figure 26]{Springel2008}.
The subhalo model is placed on an orbit constrained from the radial velocity \citep{Simon2017} and proper motion measurements \citep{Simon2018} of Tuc III. 
While recently the radial systemic velocity measurement has been refined \citep{LiTucana2018}, the peri- and apocentre of our model, $r_\mathrm{peri} \approx \unit[2.5]{kpc}$ and $r_\mathrm{apo} \approx \unit[42]{kpc}$, are roughly consistent with those tailored to match the stream \citep{Erkal2018} and given the example nature of our numerical experiments an exact match of the orbit should not be of concern. 
The ratio $r_\mathrm{peri}/r_\mathrm{apo} \sim 0.06$ is consistent with values derived from cosmological simulations of Milky Way-like haloes: for the 50 most massive satellites at $z=0$ in the \emph{Via Lactea} simulation \citep{DiemandKuhlenMadau2007}, \citet{Lux2010} find an average value of $r_\mathrm{peri}/r_\mathrm{apo} \sim 0.2\pm0.1$. Both peri- and apocentre distance of the Tucana III orbit lie approximately one standard deviation below the average values determined by \citet{Lux2010}.
We generate initial conditions by rewinding the orbit for 7 past pericentre passages. For our choice of host halo and subhalo structural parameters, this results in a tidally stripped subhalo at $z=0$ with a velocity dispersion that is compatible with Tuc III (see section \ref{sec:tuciii}).

\pagebreak[4]
\noindent\textbf{PM-code.} The numerical integration of the subhalo evolution is carried out using the particle-mesh code \textsc{superbox} \citep{Fellhauer2000}. This code employs two grids co-moving with the subhalo of resolution $\Delta x = 2 a/128$ and $20 a/128$, centred on the density maximum, as well as a fixed grid of resolution $\unit[1]{Mpc}/128$. We choose a time-step of $\Delta t = \tc(a)/400$. For the initial simulation run, this gives $\Delta x = \unit[8]{pc}$ and $\Delta t=\unit[0.5]{Myrs}$. For convergence tests, we also run models with $N = 10^7$ at a resolution of $2a/256$ and $2a/128$.

\subsection{Controlled simulation}
\label{sec:reconstruct_sim}

We now aim to reconstruct the density cusp during the simulation. This reconstruction is based on the assumptions that
(i) the DM density profile at apocentre can be approximated by an $\{\alpha,\beta,\gamma\}$ profile (see eq. \ref{eq:betagammaprofile} ), and (ii) the central slope of that profile equals $\gamma = - \diff \ln \rho / \diff \ln r = 1$.
The cusp reconstruction (\emph{boost}) then involves the following steps:
(i) At apocentre, we fit an $\{\alpha,\beta,\gamma\}$-profile to the bound particles of the subhalo, fixing $\alpha=1$, $\gamma = 1$ and matching $\rmax$ and $\Mmax=M(<\rmax)$ of the simulated subhalo, where by $\rmax$ we denote the radius of maximum circular velocity. (ii) We then generate an equilibrium $N$-body realisation of $N=10^7$ particles of the fitted density profile as described in section \ref{sec:eqmodels}, and place it on the orbit of the simulated subhalo. (iii) The spatial resolution $\Delta x$ of the particle-mesh code and time step $\Delta t$ of the integration routine are re-scaled with the fitted $\rmax$ and mass $\Mmax$ so that $\Delta x = 2 \rmax/128$ and $\Delta t = \tc(\rmax)/400$, i.e. we preserve the numerical resolution relative to properties of the evolved subhalo. We decide to reconstruct the cusp when the mass fraction within the innermost grid cell --  assuming a cuspy profile -- equals roughly 0.5 per cent of the current total subhalo mass. We have chosen this mass scale after performing convergence tests to verify that at these scales the unresolved centre of the cusp does not alter the tidal evolution of the subhalo.
For a simulation with initial spatial resolution of $\Delta x = 2 \rmax/128$, under the assumption of \citet{dehnen1993} density profiles (where $\rmax=a$), this corresponds to a halo scale radius that has decreased by a factor of $\sim 4$ due to tidal stripping. For the simulated subhalo, this means re-constructing the density cusp approximately every three pericentre passages. 
Initial structural parameters of the cusp reconstructions are listed in Table~\ref{tab:simparam}.

Figure \ref{fig:cusprec} illustrates the $3^\mathrm{rd}$ cusp reconstruction (\emph{boost}) of our simulation, performed at the $10^\mathrm{th}$ apocentre, i.e. after 9 pericentre passages.  
The left-hand panel shows the mass per radial shell $dM/dr \propto r^2 \rho(r)$ as a function of radius $r$.
The fitted $\{\alpha,\beta,\gamma\}$-profile matches the evolved $N$-body model
(corresponding to the last apocentre snapshot of the simulation starting from the $2^\mathrm{nd}$ cusp reconstruction)
well at the radii where most particles are located, i.e. around the radius $r_{-2}$ where $\diff \ln \rho / \diff \ln r|_{r_{-2}} = -2$ and the $r^2 \rho(r)$ curve peaks. For $r < \Delta x$, the fitted profile re-constructs the density cusp.
At apocentre, the subhalo is surrounded by \emph{extra-tidal} material with energies close to zero as a result of past pericentre passages \citep[e.g.][]{Penarrubia2009}. For the evolved $N$-body model in Figure \ref{fig:cusprec}, this extra-tidal material is visible for radii larger than a few scale radii. The extra-tidal material is not in equilibrium with the subhalo. To compare the differential energy distribution $n(E)$ and \DF\ of our reconstruction to the evolved $N$-body model, we allow the evolved $N$-body model to relax in isolation for 4 dynamical times. Both the differential energy distribution $n(E)$ and \DF\ of this relaxed model are well matched by the reconstruction except for energies close to zero. Cusp reconstructions consequently do not conserve the total mass or total energy of the subhalo. To give a numerical example, while $\Mmax$ differs between the final snapshot of the simulation starting from the $2^\mathrm{nd}$ cusp reconstruction and the first snapshot of the $3^\mathrm{rd}$ cusp reconstruction by $\approx\unit[1]{\mbox{per cent}}$, the total binding energy between the snapshots differs by $\approx\unit[10]{\mbox{per cent}}$ - driven by the large fraction of mass at energies close to zero, where the fit matches the $N$-body model less accurately. Our convergence tests however show that the tidal evolution is insensitive to a precise match at these energies, or similarly, to a precise match of the subhalo outer profile.

\begin{figure}
 \centering
 \includegraphics{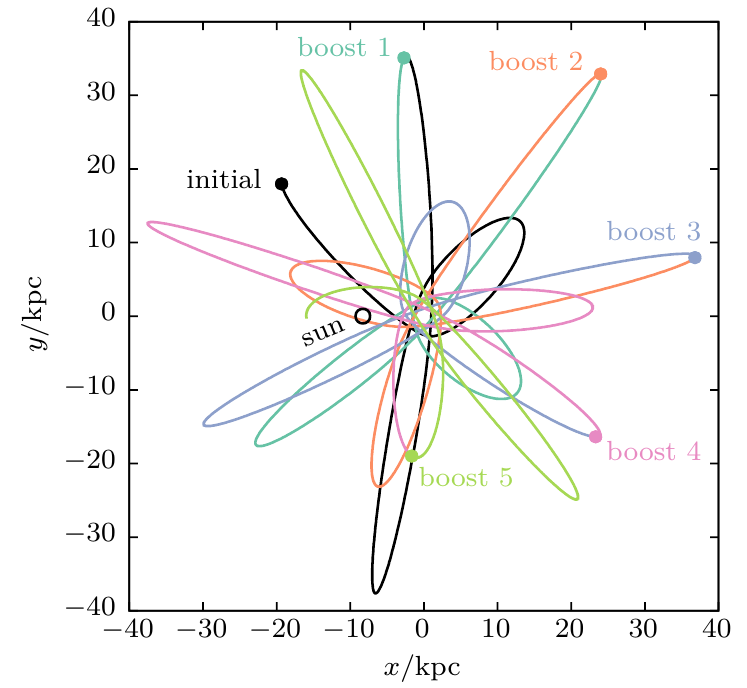}
 \caption{Projection of the subhalo orbit on the galactic plane, with orbital parameters chosen to approximate those of the Tucana III dwarf galaxy ($r_\mathrm{peri} \approx \unit[2.5]{kpc}$, $r_\mathrm{apo} \approx \unit[42]{kpc}$). We periodically reconstruct the central density cusp of the subhalo and adapt the resolution of the simulation to the subhalo structural parameters. These \emph{boosts} are shown using different colours, with the corresponding first apocentre indicated by a filled circle. The current ($t=0$) position of the sun $(x_\odot,y_\odot)=(\unit[-8.29]{kpc},0)$ is marked by an open circle.   }
 \label{fig:orbitplot}
\end{figure}

\begin{figure}
 \centering
 \includegraphics{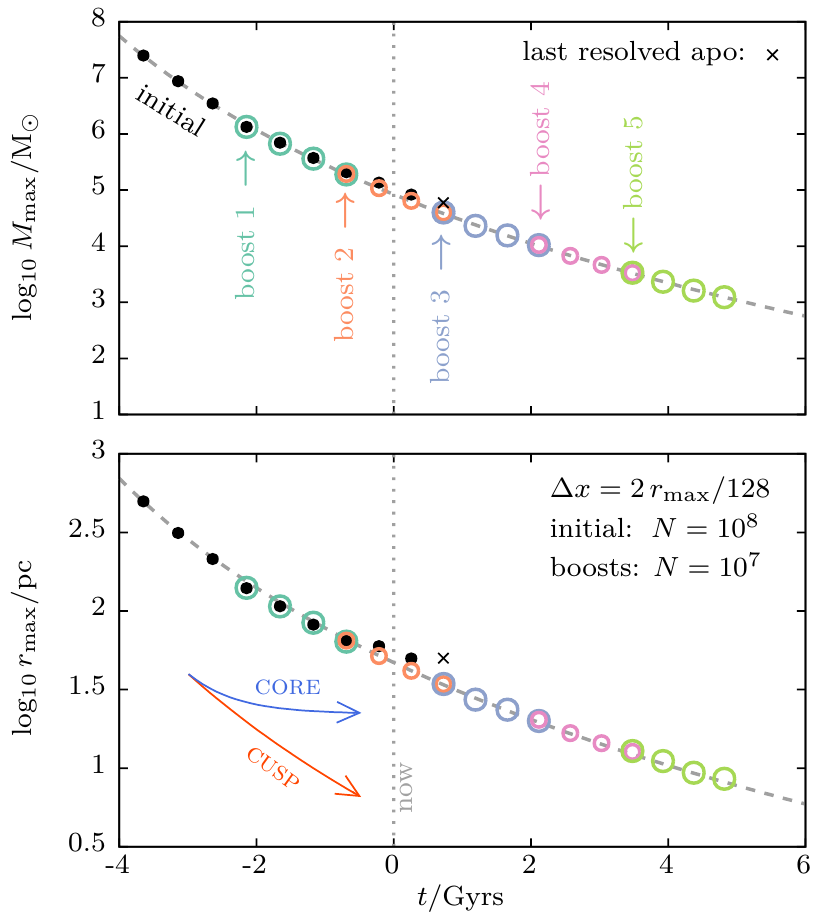}
 \caption{Evolution of $\Mmax = M(<\rmax)$ (top panel) and $\rmax$ (bottom panel) of a subhalo with initial total mass $M = \unit{10^8}{\Msol}$ and scale $a = \unit[0.5]{kpc}$. Each point corresponds to an apocentre passage of the subhalo. Black filled circles mark the simulation with highest resolution, a cross indicates the last apocentre before (artificial) disruption. Coloured circles are simulation runs (\emph{boosts}) where the density cusp is reconstructed as detailed in section \ref{sec:reconstruct_sim}: by reconstructing the density cusp, the subhalo does not disrupt, and can be traced over many orders of magnitude in mass loss. The evolution of $\rmax$ is highly sensitive to the spatial resolution of the simulation and \emph{artificially} flattens off once the simulation fails to resolve the subhalo peak circular velocity. Tidal evolutionary tracks \citep{EPW18} for cuspy and cored subhaloes are shown as a reference. Dashed curves show fits of equation \ref{eq:massevol} to the mass evolution (top panel) and the predicted $\rmax$ evolution assuming power-law scaling $\rmax \propto \Mmax^{\kappa}$ (bottom panel).}
 \label{fig:boostDM}
\end{figure}

The orbit of the simulated subhalo is shown in Figure \ref{fig:orbitplot}, and the apocentres where we reconstruct the density cusp, adapt spatial resolution $\Delta x$ and time step $\Delta t$ to the evolved structural parameters and re-start the simulation (\emph{boosts}) are marked by filled points.
Figure \ref{fig:boostDM} shows the tidal evolution of $\Mmax$ and $\rmax$ of the subhalo, measured at subsequent apocentres: by periodically reconstructing the density cusp, we can follow the tidal evolution of the subhalo for arbitrarily large fractions of tidally stripped mass.
We choose to show $\{\Mmax,\rmax\}$ instead of $\{M,a\}$ as the former can be computed directly from the $N$-body data without further assumptions about the DM profile shape. The \emph{boosts} (with $N=10^7$) follow closely the evolution of the highest-resolving \emph{initial} simulation (with $N=10^8$) for those apocentre snapshots with $\rmax \ll \Delta x$, where by $\Delta x$ we denote the spatial resolution of the particle mesh.

The last apocentre snapshot of the highest-resolving \emph{initial} simulation is indicated by a cross in Figure \ref{fig:boostDM}: beyond this snapshot, the radius of maximum circular velocity $\rmax$ is not resolved by the simulation. 
\citet{Power2003} argues that subhaloes are resolved for those radii $r$ where the acceleration $a(r)=GM(<r)/r^2$ does not exceed a \emph{characteristic} acceleration, which depends on the gravitational force softening length $\epsilon$. Based on this idea, \citet{vdBOgiya2018} propose that a subhalo can be considered sufficiently resolved if $r_\mathrm{h}/\epsilon > 7^{+3}_{-2}$, where $r_\mathrm{h}$ indicates the half-mass radius of the subhalo. For \citet{dehnen1993} profiles, and assuming that $\epsilon \approx \Delta x = 2 r_\mathrm{max,0}/128$, this translates to $r_\mathrm{max,0} / \rmax < 22^{+8}_{-6}$, where by $r_\mathrm{max,0}$ we denote the radius of maximum circular velocity at the beginning of the simulation. The upper limit of this criterion when applied to the highest-resolving \emph{initial} simulation corresponds to $\declog(\rmax/\mathrm{pc}) \approx 1.5$, of the same order as the value measured for the last resolved snapshot, $\declog(\rmax/\mathrm{pc}) \approx 1.7$.

The mass evolution is well fitted using the model of \citet{vdBTormen2005} who postulate an orbit-averaged mass loss rate of $\diff M / \diff t = - M \psi^\zeta / \tau$, where $M$ denotes the subhalo mass, $\psi = M/M_\mathrm{host}$ and $\zeta$ is a constant. In the following, we will use $\Mmax$ as a proxy for subhalo mass, and the host halo scale mass as measure for $M_\mathrm{host}$. As our numerical experiment covers only a narrow range of redshifts, we will assume in the following a static host mass: during the simulated redshift interval of $\Delta z \approx 0.6$ between the first and the last snapshot shown in Figure \ref{fig:boostDM}, the host halo scale mass increases by a factor $\exp(2a_g\Delta z)$ (see section \ref{sec:numsetup}), i.e. by less than 30 per cent, whereas the subhalo mass decreases by more than four orders of magnitude. 
Integration yields
\begin{equation}
 \Mmax(t) =
\begin{cases} 
 M_\mathrm{max,0} ~\exp(-t/\tau) & ~~\mathrm{if}~\zeta = 0 \\[5pt]
 M_\mathrm{max,0} ~\left[ 1 + \zeta \psi_0^\zeta (t/\tau)  \right]^{-1/\zeta}  & ~~\mathrm{if}~\zeta \neq 0 
\end{cases}~~~~,
 \label{eq:massevol}
\end{equation}
where $M_\mathrm{max,0} = M_\mathrm{max}(t=0)$ and $\psi_0 = {M_\mathrm{max,0}}/{M_\mathrm{host}}$. A fit of equation \ref{eq:massevol} to the simulated data with $M_\mathrm{host} = M_\mathrm{s} = \unit[1.53 \times 10^{11}]{M_\odot}$ (see section \ref{sec:numsetup}) results in a characteristic time $\tau = \unit[(160 \pm 5)]{Myrs}$ (i.e. $\tau/T_\mathrm{orb} \approx 0.34$), a power-law index for the dependence on host mass of $\zeta = 0.118 \pm 0.002$, and  $M_\mathrm{max,0} = \unit[(8.4 \pm 0.1)\times 10^4]{M_\odot}$. The fit is shown using a dashed line in the top panel of Figure \ref{fig:boostDM}. Note that this fit serves to parametrize the mass loss of a specific subhalo, whereas average mass loss rates for the entire population of subhaloes of a given merger are generally lower \citep{Giocoli2008,JiangVDB2016}. 

The slight departure of the mass-evolution from an exponential ($\zeta \neq 0$) may be tied to the non-self similar evolution of the subhalo, i.e. the steepening of the outer slope $\beta$ of the density profile during tidal evolution. The rate $\diff \Mmax / \diff t$ of tidal stripping is largest when the subhalo model is first injected in the host potential. We generate equilibrium realisations of $N$-body models in isolation, and in the case of the \citet{dehnen1993} profile with $\{\alpha,\beta,\gamma\}=\{1,4,1\}$, the differential energy distribution is a monotonously increasing function for $E \rightarrow 0$ (see central panel of Figure \ref{fig:eqcore}), i.e. there is a large fraction of particles at low binding energies. Those $N$-body particles at low binding energies are stripped once the subhalo is injected. With the steepening of the outer slope $\beta$ during tidal stripping ($\beta=4$ at infall, $\beta=7$ after 9 pericentre passages), a smaller mass fraction of the subhalo is associated to low binding energies: for the $\{1,7,1\}$ profile, $n(E) \rightarrow 0$ for $E \rightarrow 0$ (central panel of Figure \ref{fig:cusprec}), and the subhalo becomes more resilient to tides. This observation may be inverted: as self-bound (sub)haloes in Once $\rmax$ is of the same order as the resolution $\Delta x$,an external potential cannot have particles with energies arbitrarily close to zero, assuming profiles with isotropic velocity dispersion, their density must decrease more rapidly than $\mathcal{O}(r^{-4})$.

The radius $\rmax$ of maximum circular velocity decreases during tidal stripping, as shown in the bottom panel of Figure \ref{fig:boostDM}. Once $\rmax$ is of the same order of magnitude as the resolution $\Delta x$, the evolution of $\rmax$ flattens off, i.e. $\diff \rmax /\diff t \rightarrow 0$. This behaviour is symptomatic of the formation of the artificial density core as it resembles that of cored DM subhaloes in controlled collisionless simulations \citep{EPW18} -- tidal evolutionary tracks for cuspy and cored systems are plotted with solid lines Figure \ref{fig:boostDM}, using the smooth mass evolution of equation \ref{eq:massevol}.

Furthermore, we fit a power-law $\rmax \propto \Mmax^{\kappa}$ to the subhalo mass-size evolution. The fitted slope $\kappa = 0.415 \pm 0.001$ is lower than the value found in \citet{EPW18} from an average of re-simulations of the Aquarius A2 merger tree ($\kappa\approx0.48$). This may be related to strong disc shocking of our subhalo model experienced due to the particularly low pericentre distance ($r_\mathrm{peri} \approx \unit[2.5]{kpc}$), potentially heating up the subhalo, affecting its mass-size evolution.
The dashed line in the bottom panel of Figure \ref{fig:boostDM} shows the size evolution as described through the power-law fit in combination with the mass evolution of equation \ref{eq:massevol}.

In agreement with the cuspy model of section \ref{sec:tdyn}, the period $\tc(\rmax)$ of a circular orbit with radius $\rmax$ decreases during tidal stripping: assuming $\rmax \propto \Mmax^{\kappa}$, for $\diff \Mmax / \diff t < 0$, we find $\diff  \tc(\rmax) / \diff t < 0$ if $\kappa > 1/3$. This is satisfied by the fitted value of $\kappa$ for the simulated cuspy subhalo. The subhalo therefore has increasing multiples of its dynamical time to relax and reach equilibrium between subsequent tidal interactions. In contrast, $\kappa \rightarrow 0$ for cored subhaloes as the $\rmax$ evolution flattens off \citep{EPW18}, and $\tc(\rmax)$ increases during tidal stripping.

\begin{table}
\centering
\small
\renewcommand{\arraystretch}{1.2}
  \caption{Initial structural parameters of DM subhaloes and stellar populations used in the simulation runs (see sections \ref{sec:reconstruct_sim} and \ref{sec:tuciii}). Simulations are started at time $t$, where times $t>0$ lie in the future. The table lists the enclosed mass $\Mmax$ within the radius of maximum circular velocity $\rmax$ and outer slope $\beta$ for $\{\alpha,\beta,\gamma\}=\{1,\beta,1\}$ DM profiles, as well as luminosity $L$, half-light radius $\Rh$ and dynamical mass-to-light ratio $\langle M/L \rangle$ (averaged within $\Rh$) for the embedded stellar Plummer spheres.}
  \begin{tabular}{lr@{$~~~~~$}c@{$~~~~~$}c@{$~~~~~$}c@{$~~~~~$}c@{$~~~~~$}c@{$~~~~~$}c@{$~$}}
  \toprule
          & $\displaystyle\frac{t}{\mathrm{Gyrs}}$ & $\displaystyle\frac{\Mmax}{\unit{M_\odot}}~~$ & $\displaystyle\frac{\rmax}{\unit{pc}}$ & $\beta~$ & $\displaystyle\frac{L}{\unit{L_\odot}}$  & $\displaystyle\frac{\Rh}{\unit{pc}}$ & $\displaystyle\frac{\langle M/L \rangle}{ \unit{M_\odot / L_\odot}}$ \\ \cmidrule(l){2-8}
  initial & $-3.6~$ & $2.5 \times 10^7$      & 500               & 4.0     & $3100$              & $35$         & 270     \\
  boost 1 & $-2.1~$ & $1.3 \times 10^6$      & 140               & 5.5     & $2600$              & $44$         & 220     \\
  boost 2 & $-0.7~$ & $1.9 \times 10^5$      & 64                & 6.0     & $1400$              & $50$         & 200     \\
  boost 3 & $0.7~$  & $4.0 \times 10^4$      & 34                & 7.0     & $230$               & $41$         & 410     \\
  boost 4 & $2.1~$  & $1.0 \times 10^4$      & 20                & 7.5     & $18$                & $29$         & 1600     \\
  boost 5 & $3.5~$  & $3.2 \times 10^3$      & 12                & 8.0     & $1.2$               & $20$         & 8200    \\
  \bottomrule

  \end{tabular}
  \label{tab:simparam}
\end{table}

\section{Application to Milky Way dSphs: Tucana III}
\label{sec:tuciii}

We now apply the method of reconstructing the density cusp to study the tidal evolution of dwarf galaxies embedded in cuspy DM subhaloes. This is of particular interest because of the recent discoveries of several faint, low-mass dwarf galaxies in the Milky Way \citep[e.g.][]{Drlica-Wagner2015,Koposov2015,Torrealba2016}, some of them showing tidal features. 
Motivated by the large dynamical mass-to-light ratios inferred for Milky Way dwarf galaxies \citep[e.g.][]{Walker2007}, the following analysis is carried out under the assumption that stars are mass-less tracers of the underlying DM potential. This allows us to model the evolution of the stellar component using the \DF-based method introduced by \citet{Bullock2005}: 
If both the stellar and DM density distributions are spherical, assuming isotropic velocity dispersion profiles, their {\DF}s can be written as functions of energy $E$. Then, in the notation of equation \ref{eq:diffenergy}, 
the probability of an $N$-body particle with energy $E$ to represent a star is proportional to 
\begin{equation}
 \mathcal{P}_\star(E) \propto \frac{n_\star(E)}{n(E)} = \frac{f_\star(E)}{f(E)}~,
\end{equation}
as the potential is sourced by DM only and therefore the density of states $p_\star(E) = p(E)$ cancels.  We compute the probabilities $\mathcal{P}_\star(E)$ at infall, and re-compute them after each cusp reconstruction. Structural and kinematic properties of the stellar component can be inferred from the DM distribution by applying the individual $\mathcal{P}_\star(E)$ as weights. A basic implementation of this method is made available online together with code to generate equilibrium $N$-body models, see section \ref{sec:eqmodels}.

Luminosity and projected half-light radius of the stellar component are chosen so that the evolved model approximately matches the Tucana III dwarf galaxy at redshift $z=0$: $L = \unit[780^{+350}_{-240}]{L_\odot}$, $\Rh = \unit[(44\pm6)]{pc}$ \citep{Drlica-Wagner2015}, with an upper limit on the line-of-sight velocity dispersion of $\sigma < \unit[1.5]{\kms}$ \citep{Simon2017}. Using the upper limit on $\sigma$, these authors find an upper limit on the mass enclosed within the half-light radius of $M(<\Rh) < \unit[9 \times 10^4]{M_\odot}$ and an upper limit on the dynamical mass-to-light ratio averaged within $\Rh$ of $\langle M/L \rangle = 2 M(<\Rh) / L < \unit[2.4 \times 10^2]{ M^{\phantom{1}}_\odot L^{-1}_\odot}$.
We model the stellar density profile to closely resemble a Plummer sphere, i.e. $\{ \alpha_\star, \beta_\star, \gamma_\star \} = \{ 2, 5, 0 \}$ in the notation of equation \ref{eq:betagammaprofile}. Note that a strictly cored ($\gamma_\star=0$) stellar tracer profile cannot be embedded self-consistently in equilibrium in a cuspy ($\gamma=1$) DM halo: Eddington inversion for such tracer - potential pairs results in {\DF}s that do not satisfy $f(E)\geq0$ for all energies $E$. 
In Appendix \ref{appendix:gammamin} we compute the minimum central slope $\gamma_{\star,\mathrm{min}}$ of a collisionless stellar tracer embedded in a cuspy DM halo. Interestingly, the minimum slope $\gamma_{\star,\mathrm{min}}$ increases with the ratio between stellar and DM scale radius. As tidal stripping tends to increase this ratio, assuming that the stellar tracer remains spherical and isotropic, a consequence of tidal evolution is the formation of a shallow density cusp in the stellar tracer profile. 

\begin{figure}
 \centering
 \includegraphics{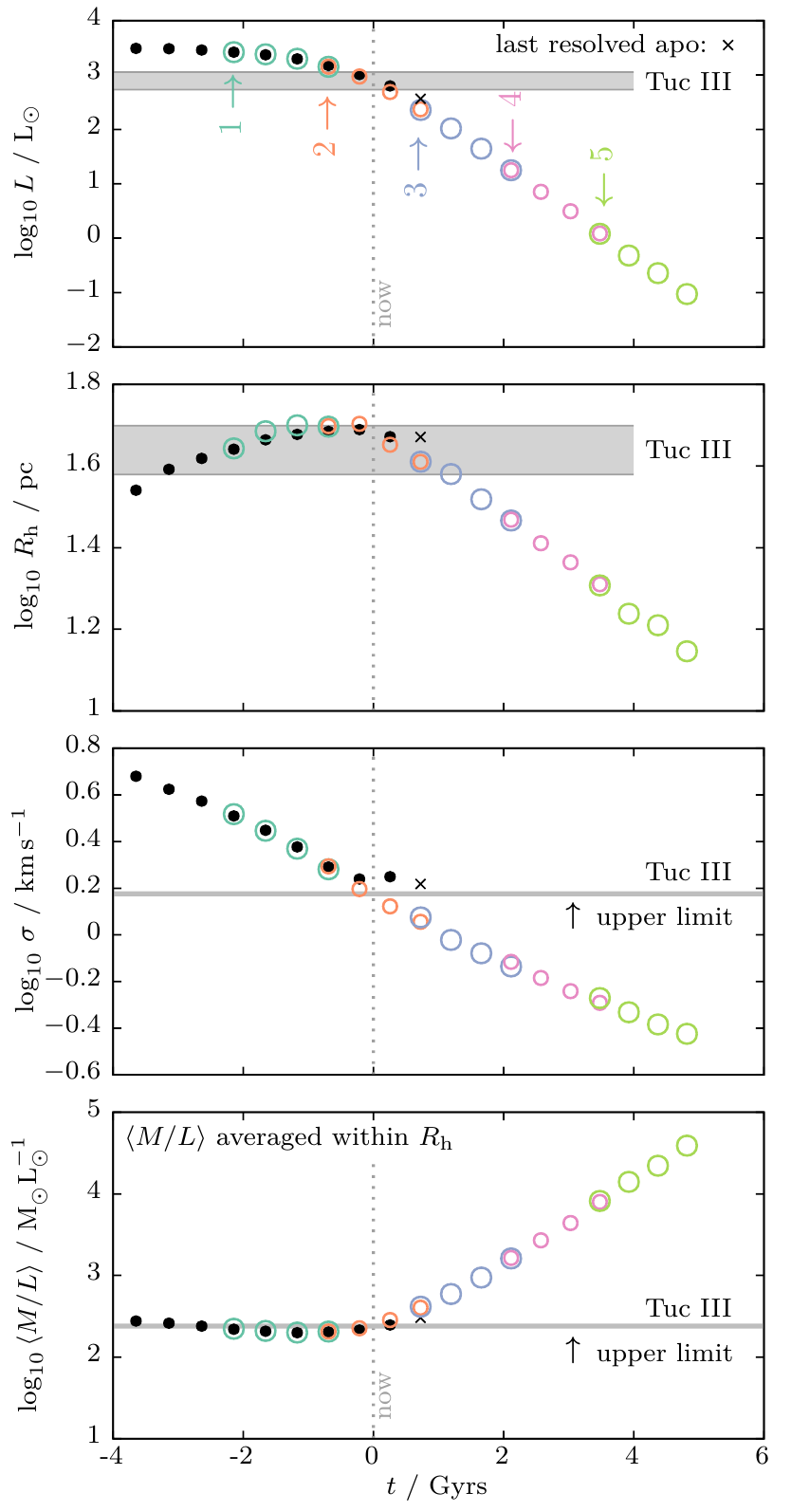}
 \caption{Evolution of luminosity $L$, half-light radius $\Rh$, luminosity-averaged line-of-sight velocity dispersion $\sigma$ and dynamical mass-to-light ratio $\langle M/L \rangle$ (averaged within $\Rh$) of a dwarf spheroidal galaxy embedded in the cuspy DM subhalo with parameters as listed in section \ref{sec:numsetup}. The values of $L$, $\Rh$ and $\sigma$ at $z=0$ are chosen to approximate the observed properties of the Tucana III dwarf. By periodically reconstructing the density cusp of the underlying DM subhalo, we can follow the evolution of the embedded dwarf galaxy down to sub-solar luminosities: the dwarf galaxy is not disrupted by tides.   
 Different colours correspond to cusp reconstruction \emph{boosts} as in Figure~\ref{fig:boostDM}. }
  \label{fig:booststars}
\end{figure}

At each reconstruction of the DM density cusp, we fit a Plummer profile (allowing for a shallow density cusp, $\gamma_\star \lesssim 0.1$) to the stellar component and calculate the stellar probabilities $\mathcal{P}_\star(E)$ to match the fitted profile. Structural parameters for the embedded stellar profiles are listed in Table \ref{tab:simparam}.
Figure \ref{fig:booststars} shows the evolution of the stellar component for the same apocentre snapshots of the DM subhalo in Figure \ref{fig:boostDM}, distinguishing cusp reconstruction \emph{boosts} using different colours.
While initially the dwarf galaxy loses predominantly DM and its luminosity $L$ decreases only marginally, once the stellar half-light radius $\Rh$ is of the same order as the subhalo $\rmax$, stars get stripped efficiently. This is consistent with the findings of previous studies on the evolution of cored stellar tracers embedded in cuspy DM subhaloes using controlled simulations with non-adaptive resolution \citep[e.g.][]{penarrubia2008,EPT15}.

Grey shaded stripes in Figure \ref{fig:booststars} indicate the measured luminosity $L$ and half-light radius $\Rh$ of the Tucana III dwarf respectively, whereas a grey solid line marks the upper limit on the measured velocity dispersion $\sigma$ and dynamical mass-to-light ratio $\langle M/L \rangle$. Our $N$-body model is consistent with these observables at $z=0$. Note that the dynamical mass-to-light ratio $\langle M/L \rangle$ increases during tidal evolution, reaching values as extreme as $\sim 10^4$: our assumption of stars being collisionless tracers of the underlying potential therefore holds for the modelled evolution of Tuc III.
We follow the tidal evolution of the Tucana III model down to sub-solar luminosities and a luminosity-averaged line-of-sight velocity dispersion of $\sigma = \langle \sigma_\mathrm{los}^2 \rangle^{1/2} < \unit[0.5]{km\,s^{-1}}$. Similar to the evolution of the underlying DM subhalo, the embedded dwarf galaxy is not disrupted by tides. Abundance and detectability of such highly stripped dwarf galaxies will be discussed in the following section.

\section{Summary and discussion}
\label{sec:discuss}

In the present work, we argue that cold dark matter subhaloes with centrally-divergent density cusps cannot be disrupted by smooth tidal forces. There are two driving causes for the tidal survival of subhaloes: (i) For circular orbits within cuspy subhaloes, the orbital period $t_\mathrm{c}(r) \rightarrow 0$ for $r \rightarrow 0$. As a consequence there is a fraction of particles within the subhalo that reacts adiabatically to perturbations by tides. Using empirical formulae for the tidal evolution of structural parameters of subhaloes obtained from controlled simulations, we show that the fraction of particles that react adiabatically increases during tidal evolution in cuspy haloes. (ii) Furthermore with $t_\mathrm{c}(r/a)$ decreasing during tidal evolution, the subhalo has increasing multiples of its dynamical time to relax and reach equilibrium between subsequent tidal interactions.
On the other hand, for subhaloes with constant-density cores, $t_\mathrm{c}(r) \rightarrow \mathrm{const}>0$ for $r \rightarrow 0$. Tidal evolution decreases the fraction of particles that react adiabatically to tidal perturbations in cored haloes, and with dynamical times increasing during tidal evolution, it becomes increasingly difficult for cored subhaloes to relax and reach equilibrium between subsequent tidal interactions. This facilitates the tidal disruption of cored subhaloes. 

Using controlled simulations, in the spirit of \citet{vdb2018}, we show how insufficient numerical resolution causes the \emph{artificial} formation of constant-density cores in collisionless numerical simulations of initially cuspy subhaloes. Under the assumption that tides do not alter the central slope $\gamma = - \diff \ln \rho / \diff \ln r = 1$ of cuspy subhaloes \citep[e.g.][]{Hayashi2003,Penarrubia2010}, we perform a numerical experiment where we periodically reconstruct the density cusp of a subhalo evolving in a Milky Way-like potential. This prevents the artificial disruption of the subhalo due to insufficient numerical resolution and allows us to follow its evolution for arbitrarily large fractions of tidally stripped mass.  We furthermore study the evolution of dwarf galaxies embedded in cuspy DM haloes under the assumption that stars are collisionless tracers of the underlying potential. Using a model of the Tucana III dwarf as an example, we show that dwarf galaxies embedded in cuspy haloes can be stripped to sub-solar luminosity by tides.

\subsection{Limitations of the model}
Several aspects of our numerical experiments call for caution when drawing quantitative conclusions about the physical universe, though none of these limitations effect our main conclusion of the tidal survival of cuspy DM subhaloes. 
(i) We modelled DM subhaloes as $N$-body realisation with isotropic velocity dispersion profiles. Cosmological simulations indicate that the central regions of DM haloes do have isotropic velocity dispersions \citep{NavarroLudlow2010,KlypinMultiDark2016}, and we have verified that our tidally stripped subhalo models remain isotropic in the centre.
(ii) When reconstructing the density cusp, we neglect the effect of extra-tidal material on the subsequent evolution of the subhalo. 
While extra tidal material has an effect of dynamical friction on the subhalo \citep{Fujii2006,Fellhauer2007,vdBOgiya2018}, we find that extra tidal material does not alter the tidal evolution of the subhalo: reconstructions tailored to match the subhalo potential sourced by bound particles evolve in agreement with the original model. This is consistent with the results of controlled simulations by \citet{vdBOgiya2018}, who find that for an initial ratio $M/M_\mathrm{host} \sim 1/1000$ of subhalo mass $M$ and host mass $M_\mathrm{host}$, dynamical friction from tidally stripped material affects the orbital radius by a few per cent at most.
(iii) Our subhalo models are strictly collisionless. The importance of collisionality depends on the number of particles that make up the subhalo, and therefore on the mass of the smallest DM constituents, which may range from earth mass for neutralinos \citep[e.g.][]{DiemandMoore2005} to the order of (multiple) solar masses for primordial black holes \citep[e.g.][]{Bird2016}. 
(iv) We furthermore model stars as massless tracers of the underlying potential. This assumption is well motivated by the large dynamical mass-to-light ratios of our models, $10^2 \lesssim \langle M/L \rangle / (M_\odot/L_\odot) \lesssim 10^4$ (averaged within the half-light radius, see Figure \ref{fig:booststars}). 
Baryons embedded in DM haloes may also alter the DM halo profile through feedback \citep[e.g.][]{Pontzen2012}, causing density cores which can be tidally disrupted -- though DM cusps may reform after baryonic feedback eases \citep{LaportePenarrubia2015}.
(v) We only considered the effect of smooth tidal fields. Substructures present in the host halo, e.g. other DM subhaloes, giant molecular clouds and stars, may significantly heat up a subhalo and increase the rate of tidal stripping \citep{Penarrubia2019, Delos2019}, although adiabatic response in inner regions may prevent full disruption \citep[e.g.][]{Weinberg1994}.

\begin{figure}
 \centering
 \includegraphics{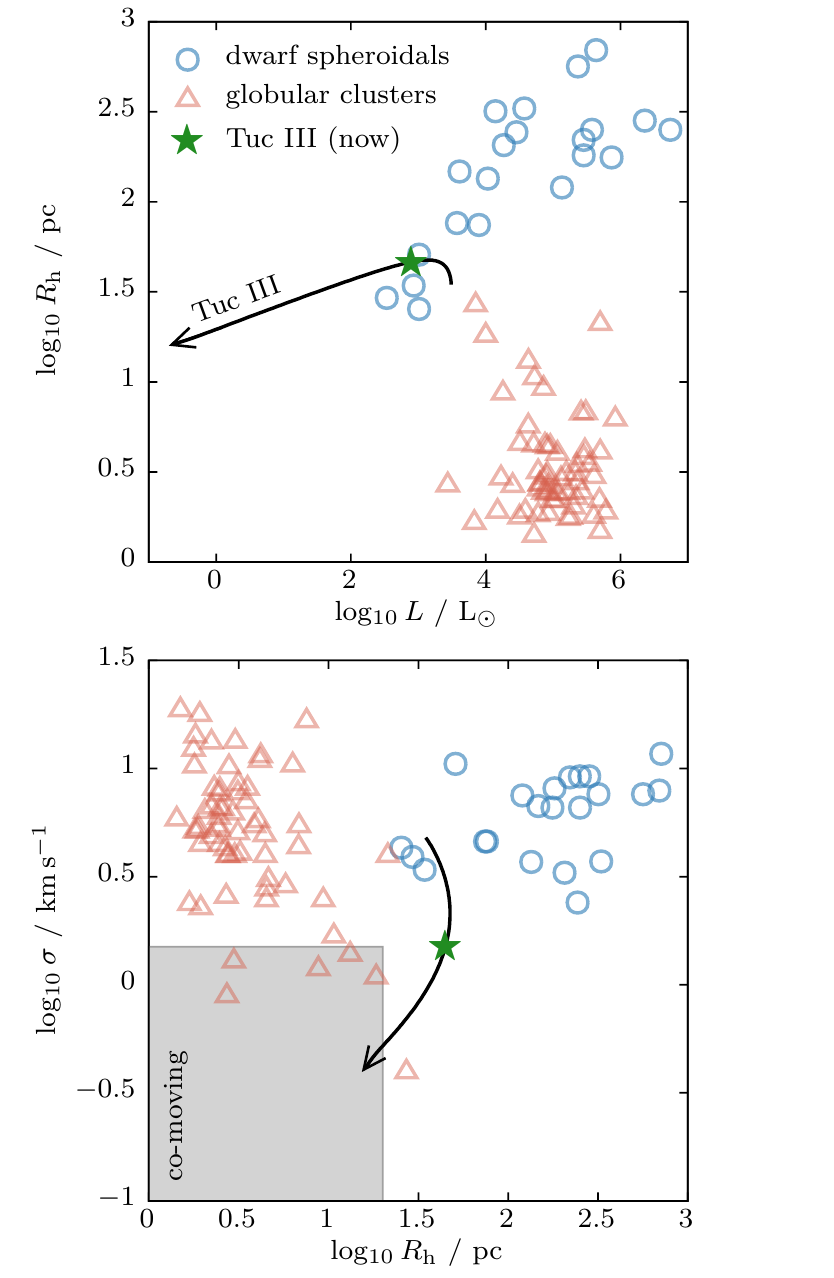}
 \caption{Evolution of the Tuc III model compared to luminosities $L$, half-light radii $\Rh$ and velocity dispersions $\sigma$ of a sample of Milky Way dwarf galaxies (from \citealt[][]{McConnachie2012} with the additions as listed in \citealt[table 2]{EPW18}  ) and globular clusters \citep[from][2010 revision]{Harris1996}. An evolved dwarf remnant would appear as a co-moving group of stars with a half-light radius compatible with those of globular clusters but much lower luminosity and velocity dispersion. The grey shaded area in the bottom panel marks the range of stellar separations and dispersions of co-moving stellar pairs that are likely to have formed together \citep{Kamdar2019}.}
  \label{fig:gcdsph}
\end{figure}

\begin{figure}
 \centering
 \includegraphics{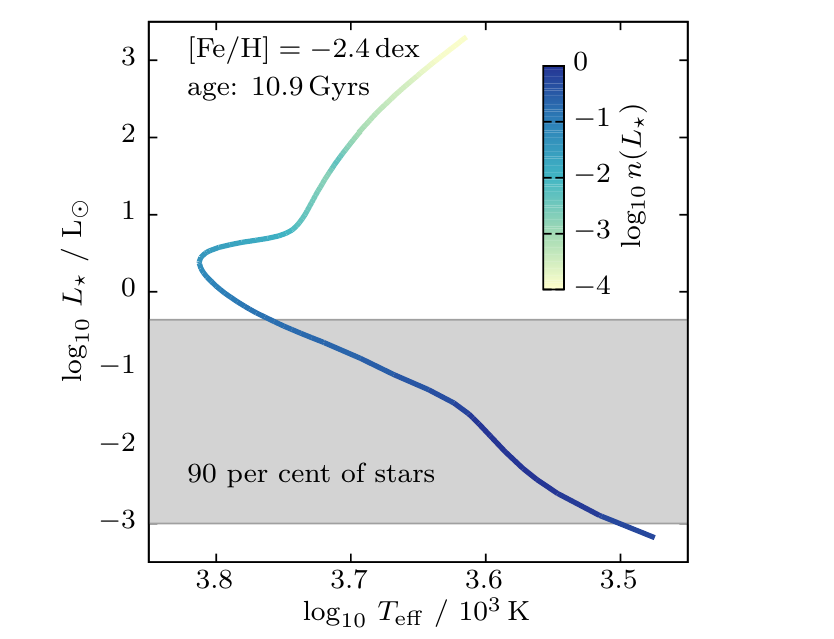}
 \caption{\textsc{parsec} isochrone \citep{Bressan2012} for a metallicity of $\unit{[Fe/H]}=\unit[-2.4]{dex}$ and an age of \unit[10.9]{Gyrs}, approximating the stellar population of the Tucana III dwarf \citep{Drlica-Wagner2015, Simon2017}. The relative abundance $n(L_{\star})$ of stars with luminosity $L_\star$ is colour-coded, highlighting that 90 per cent of stars are located below the main-sequence turnoff. }
  \label{fig:cmd}
\end{figure}

\subsection{Detectability of stripped subhaloes and dwarf remnants}
The survival of low-mass DM subhaloes has implications for potential detection through annihilation signals \citep[e.g.][]{Lavalle2007,Stref2019}, strong gravitational lensing \citep[e.g.][]{Vegetti2009,Despali2017}, pulsar-timing arrays \citep[][]{Kashiyama2018, Dror2019}, the number of gaps to be expected in stellar tidal streams \citep[e.g.][]{Ibata2002heating, ErkalBelokurov2014}, and the stochastic tidal heating of gravitating substructures \citep{Penarrubia2019}.

A detailed estimate of the abundance of subhaloes stripped to sub-kpc scale lies beyond the reach of the numerical experiments discussed in the present work. 
However it is possible to estimate the abundance of potential progenitors to micro-galaxies with sub-solar luminosities from the re-simulations of the Aquarius A2 merger tree introduced in \citet{EPLG17}. Those re-simulations follow the tidal evolution of the $\sim 10^3$ cuspy subhaloes which at the peak of their mass evolution reached a mass $M \geq \unit[10^8]{M_\odot}$, sufficiently massive to allow star formation \citep{Gnedin2000}. In presence of a galactic disc, at redshift $z=0$, of the order of $\sim 200$ subhaloes were stripped to masses below the resolution limit of the re-simulation $M \lesssim \unit[5 \times 10^5]{M_\odot}$. 
In light of the results of this paper, these subhalos may host bound visible remnants, and may therefore constitute progenitors to micro-galaxies.

\emph{How can dwarf galaxy remnants be distinguished from other clusters of stars?} 
The tidal evolution of our model of the Tucana III dwarf galaxy results in a co-moving group of stars of sub-solar total luminosity, embedded in a cuspy DM halo. Its structural properties evolve away from those of classical dwarf galaxies and globular clusters (see Figure \ref{fig:gcdsph}).
Neglecting effects off mass segregation, dwarf galaxy remnants with sub-solar luminosity will have been stripped of all their more massive (i.e. less numerous) stars, with most stars populating the low-luminosity tail of the main sequence. A \textsc{parsec} isochrone \citep{Bressan2012} for a metallicity of $\unit{[Fe/H]}=\unit[-2.4]{dex}$ and an age of \unit[10.9]{Gyrs} \citep{Drlica-Wagner2015, Simon2017}, approximating the stellar population of the Tuc III, is shown in Figure \ref{fig:cmd}. The relative abundance of stars is colour-coded, highlighting that 90 per cent of stars are located below the main-sequence turnoff. At a mean luminosity per star of $\langle L_\star \rangle/\mathrm{L_\odot}\approx0.25$, co-moving groups of total sub-solar luminosity may contain only a handful of stars.

Recently, \citet{Kamdar2019} have shown using simulations that (pairs of) stars with separations of $\unit[\Delta r<20]{pc}$ and $\unit[\Delta v<1.5]{\kms}$ are likely to have formed together, and identified 111 co-moving pairs using \emph{Gaia} data. They conclude that co-moving stars originate preferentially from star clusters younger than $\unit[1]{Gyr}$ -- such co-moving pairs should therefore have notably higher metallicities than dwarf galaxy remnants.
With large dynamical mass-to-light ratios of $10^3 \sim 10^4\,\unit{M_\odot/L_\odot}$ predicted for dwarf galaxy remnants, accurate kinematics for such systems 
would allow to constrain the presence of a DM subhalo surrounding a co-moving group of stars. 
Given the predicted low velocity dispersions $\sigma \lesssim \unit[1]{\kms}$, below the dispersion background caused by binaries \citep{McConnachieCote2010}, accurate velocity dispersions measurements may prove to be technically challenging. Seminal work by \citet{Koposov2011} however has demonstrated that accurate stellar kinematics can also be obtained for dwarf galaxies with velocity dispersions of the order of $\kms$ by repeated measurements of single-star velocities, reducing single-star velocity errors to values as low as $\unit[0.2]{\kms}$. 
Furthermore on a statistical basis, the search for extended ($\sim \unit[10^1]{pc}$) co-moving groups of stars with low metallicity and low velocity dispersion in the Milky Way halo may constitute a promising way to test the existence of \emph{micro-galaxies} predicted by CDM.

\subsection*{Acknowledgements}
\urlstyle{rm}
The authors would like to thank Jose O\~norbe, Michael Petersen and Frank van den Bosch for comments and discussions.
RE acknowledges support through the Scottish University Physics Alliance.

\footnotesize{
\bibliography{tidal}
}


\appendix
\section{Density cusps of stellar tracers embedded in cuspy dark matter haloes}
\label{appendix:gammamin}

In the present paper, we treat stars as collisionless tracers of the underlying dark matter potential. We consider potentials sourced by spherical cuspy density distributions, i.e. distributions for which $\gamma = - \diff \ln \rho / \diff \ln r  > 0$ for $r \rightarrow 0$.
We focus on the particular case of $\gamma = 1$, motivated by the \citet{Navarro96} density distribution for dark matter haloes. 
Stellar density distributions are frequently approximated by a Plummer profile, $\{\alpha_\star, \beta_\star, \gamma_\star\} = \{2,5,0\}$ in the notation of equation \ref{eq:betagammaprofile}. This is a cored density profile as $\diff \ln \rho / \diff \ln r \rightarrow 0$ for $r \rightarrow 0$. For systems with isotropic velocity dispersion, the distribution function $f(E)$ of such a (cored) stellar tracer embedded in a (cuspy) dark matter profile does not satisfy $f(E) \geq 0$ for all energies $E$: a cored stellar tracer embedded in cuspy dark matter profile cannot be realized as an equilibrium configuration.
Allowing for a shallow density cusp $\gamma_\star$ in the tracer distribution alleviates this problem.

\emph{What is the minimum central slope $\gamma_\star$ of a stellar tracer embedded in a cuspy dark matter halo?}
We address this question for spherical tracer distributions with isotropic velocity dispersion. In particular, we compute the minimum central slope $\gamma_\star$ of a stellar tracer with $\{\alpha_\star, \beta_\star, \gamma_\star\}=\{2,5,\gamma_\star\}$ density profile embedded in a dark matter halo with $\{\alpha, \beta, \gamma\}=\{1,\beta,1\}$ density profile for different outer slopes $\beta$. For this purpose, we use Eddington inversion to calculate the distribution function $f(E)$ corresponding to a given tracer and dark matter density profile (see section \ref{sec:eqmodels}). Physical distribution functions satisfy $f(E) \geq 0$ for all energies $E$.

\begin{figure}
 \centering
 \includegraphics{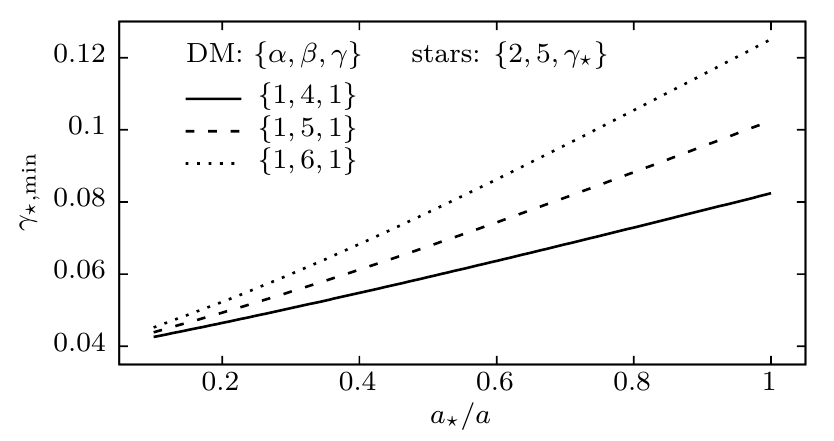}
 \caption{Minimum value $\gamma_{\star,\mathrm{min}}$ of the central slope $\gamma_\star = - \diff \ln \rho_\star / \diff \ln r$, $r\rightarrow 0$ for collisionless stellar tracers with $\{\alpha_\star,\beta_\star,\gamma_\star\}=\{2,5,\gamma_\star\}$ profile embedded in equilibrium in cuspy $\{\alpha,\beta,\gamma\}$ DM haloes. The minimum slope is shown as a function of \emph{stellar segregation}, expressed as the ratio of stellar scale radius $a_\star$ and dark matter scale radius $a$ (in the notation of equation \ref{eq:betagammaprofile}). For slopes shallower than $\gamma_{\star,\mathrm{min}}$, the distribution function for systems with isotropic velocity dispersion does not satisfy $f(E) \geq 0$ for all energies $E$. }
  \label{fig:gammamin}
\end{figure}

Figure \ref{fig:gammamin} shows the minimum central slope $\gamma_\star$ necessary to satisfy $f(E) \geq 0$ for all $E$ of a stellar tracer embedded in a cuspy dark matter halo for different choices of the outer slope $\beta=(4,5,6)$. The minimum central stellar slope $\gamma_\star$ is plotted as a function of segregation $a_\star/a$, parametrising how deeply embedded the stellar tracer distribution is within the dark matter halo, expressed as the ratio of tracer scale radius $a_\star$ and dark matter scale radius $a$. As a point of reference, note that the projected half-light radius $\Rh$ of a Plummer profile is equal to the profile scale radius $a_\star$. 
The minimum slope $\gamma_\star$ increases with the ratio $a_\star/a$ between stellar and dark matter scale radius. As tidal stripping tends to increase this ratio (see Table \ref{tab:simparam}), assuming that the stellar tracer remains spherical and isotropic, a consequence of tidal evolution is the formation of a shallow density cusp in the stellar tracer profile.

\bsp	

\label{lastpage}

\end{document}